\def\sqig{$\sim\,$} \def\etal{et\,al.} \def\msun{M$_{\scriptstyle\odot}$} 
\def\up#1{$^{\mbox{{\scriptsize #1}}}$}  
\def\lo#1{$_{\mbox{{\scriptsize #1}}}$} \def\pten#1{$\times10^{#1}$}
\def\deg{$^{\circ}$} \def\kmps{km\,s$^{-1}$} 
\def\chisq{$\chi^{2}_{\nu}$}\def\minone{$^{-1}$}\def\mintwo{$^{-2}$}
\def\sqiglt{\hbox{\rlap{\lower.55ex \hbox {$\sim$}}
	\kern-.3em \raise.4ex \hbox{$<$}\,}}
\def\sqiggt{\hbox{\rlap{\lower.55ex \hbox {$\sim$}}
	\kern-.3em \raise.4ex \hbox{$>$}\,}}
\def\kev{\,ke\kern-.10em V}
\documentstyle{mn}
\title[Accretion in intermediate polars]
{The size of the accretion region in intermediate polars: eclipses of
XY~Arietis observed with {\sl RXTE}}
\author[C. Hellier]{Coel Hellier\\
Department of Physics, Keele University, Keele, Staffordshire, ST5 5BG}
\date{Accepted 1997 May 28. Received 1997 March 13}
\begin{document}
\maketitle
\begin{abstract}
{\it RXTE\/} observed 20 eclipse egresses of the intermediate polar XY~Ari in
order to study the size and structure of the X-ray emitting accretion regions.
The spin-phase averaged egress lasts 26 s, implying a white dwarf radius of
4.3--7.0 \pten{8} cm. The individual egresses occur later in orbital
phase with later spin phase, as expected if the white dwarf spins in
the same sense as the orbital motion. The eclipse times trace out the motion of
the upper pole across the white dwarf face; then, when
the upper pole disappears over the white dwarf limb and the lower pole appears,
they trace the motion of the lower pole across the face. 

Aligning all the egresses shows that the majority of the X-ray flux emerges in
$<2$ s, implying accretion regions with area, $f$, $< 0.002$ as a fraction of
the white dwarf surface. Using only the spin-phase to align the egresses,
however, gives a longer (\sqig 5 s) time for the emergence of the majority of
the flux, implying that the accretion regions wander over an area of
$f < 0.01$. There is also evidence that a minority of the flux emerges from a
much larger area, or that we see  accretion regions at both poles
simultaneously at some spin phases. 
\end{abstract}
 
\section{Introduction}
If a close binary system is eclipsing it is far easier to obtain constraints on
the geometry of the binary. The 206-s X-ray pulsar XY~Ari (H0253+193) might
have seemed destined to obscurity by its location behind the molecular cloud
Lynds 1457, since the optical flux is extinguished to $V$\,$>$23. However,
following Patterson \&\ Halpern's (1990) suggestion that it was an intermediate
polar  (IP -- a magnetic white dwarf accreting material from a close companion),
Kamata, Tawara \&\ Koyama (1991) discovered deep X-ray eclipses recurring with
a 6.06-h orbital period. XY~Ari is now the only known IP with a deep X-ray
eclipse (EX~Hya shows a partial X-ray eclipse while DQ~Her shows a deep eclipse
but negligible X-ray flux; see Patterson 1994 for a review of these stars).

The identification of XY~Ari in the infra-red (Zuckerman \etal\ 1993) allows
modelling of the ellipsoidal variations in the light curve (Allan, Hellier \&\
Ramseyer 1996), tying down the system parameters.
Kamata \&\ Koyama (1993) made the first attempt to use the X-ray eclipse to
constrain the accretion geometry, using 5 ingresses and egresses observed with
{\it Ginga.} They concluded that the time of ingress/egress was comparable to
the expected size of the white dwarf, and showed the potential of observing
more eclipses at a higher count  rate. 

Accordingly, I proposed an observation of 20 eclipse egresses with the larger
PCA on the {\it RXTE\/} satellite (Bradt, Rothschild \&\ Swank
1993). The aim was to record changes of the eclipse egress with the phase of
the 206-s white dwarf spin cycle. I chose to observe only egresses so that they
were all directly comparable: at ingress the secondary star limb is at a
different angle on the white dwarf, potentially providing more information, 
but requiring more eclipses to look for patterns.

The motivation for the observations is that after 15 years of studying
IPs there is no agreement on the pattern of
accretion flow onto the white dwarf, with estimates of $f$, the fraction
of the white dwarf onto which accretion occurs, differing by orders of
magnitude. King \&\ Shaviv (1984), Norton \&\ Watson (1989) and Patterson 
(1994) all argue for $f >$ 0.1--0.3; while Rosen, Mason \&\ Cordova (1988),
Hellier, Cropper \&\ Mason (1991) and Hellier (1995) claim $f$ = 0.01--0.001. 
This paper presents direct geometric evidence addressing this issue. 

\begin{figure*}\vspace{8cm}   % Fig 1
\caption{One of the 20 {\it RXTE\/} lightcurves of XY~Ari, showing the star
emerging from eclipse and pulsing at the 206-s spin period.
Each bin lasts 10 s. The errors can be judged from the data in eclipse, since
the high background level means that photon noise is nearly constant.}
\includegraphics{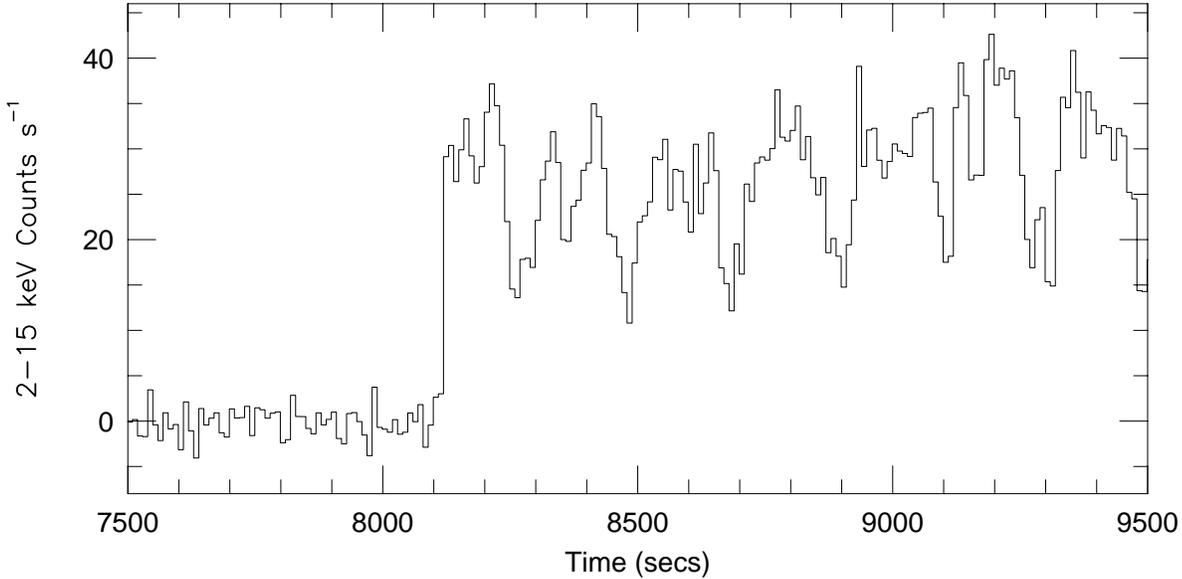}
\end{figure*}

\section{Observations}
The 20 eclipse egresses were observed by {\it RXTE\/} over the period 1996 July
13 to 1996 August 11, at a rate of roughly one per day. Each observation lasted
\sqig2000 s, with the first 1000 s in eclipse and the later half covering
\sqig5 spin cycles. As a bonus XY~Ari went into outburst during this period,
giving 14 observations in quiescence and 6 in outburst. The outburst is
discussed in a companion paper (Hellier, Mukai \&\ Beardmore 1997), while this 
paper concentrates on the eclipse.

I extracted lightcurves in the energy range 2--15 \kev, using data from the
top Xenon layer of the PCA only [this gives a higher signal-to-noise ratio
(S/N) for dim sources]. I then subtracted the background estimated using {\sc
pcabackest} v1.4f. This marginally underestimated the count rate in
eclipse, so I fitted a line to the data in eclipse and subtracted that
as a final tweak. The best lightcurve (the one with the egress occurring
at the highest count rate) is shown in Fig.~1. Further lightcurves appear
in Hellier \etal\ (1997). All times were converted to barycentric dynamical
time.

%\section{The spin cycle}
In quiescence the X-ray spin pulse of XY~Ari has a low amplitude (\sqig25
per cent) and is double-peaked (e.g.~Kamata \&\ Koyama 1993). Fig.~2 shows the
Fourier transform of the 14 quiescent observations around the first harmonic of
the  spin pulse. Since the observations are so short compared to the data gaps
the aliasing is severe, but fortunately the {\it Ginga\/} spin period of
206.298 s (Kamata \etal\ 1991) selects the correct alias and agrees well 
with the current data. 

\begin{figure}\vspace{7cm}   % Fig 2
\caption{The Fourier transform of the 14 quiescent observations of XY~Ari.
The tick mark locates the first harmonic of the 206.298-s spin pulse.}
\includegraphics{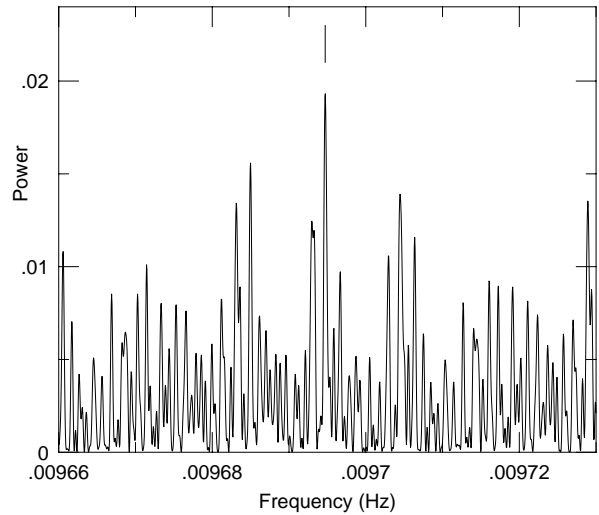}
\end{figure}

\begin{figure*}\vspace{8cm}   % Fig 3
\caption{The averged eclipse egress of XY~Ari. Each bin lasts 2 s.
Third contact occurs at TDB 245\,0278.82050(1).}
\includegraphics{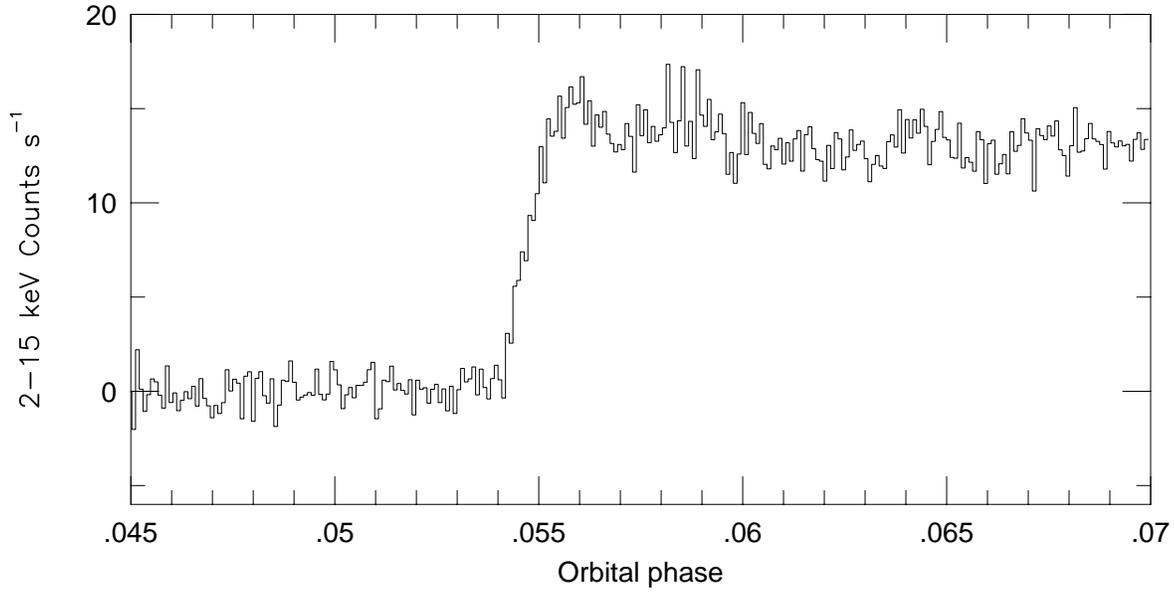}
\end{figure*}

\begin{figure*}\vspace{12cm}   % Fig 4
\caption{Three of the 20 egresses, shown at 1-s resolution. The uppermost
egress is that from Fig.~1, shown at a higher time resolution. As can be
seen, the egresses occur at different orbital phases.}
\includegraphics{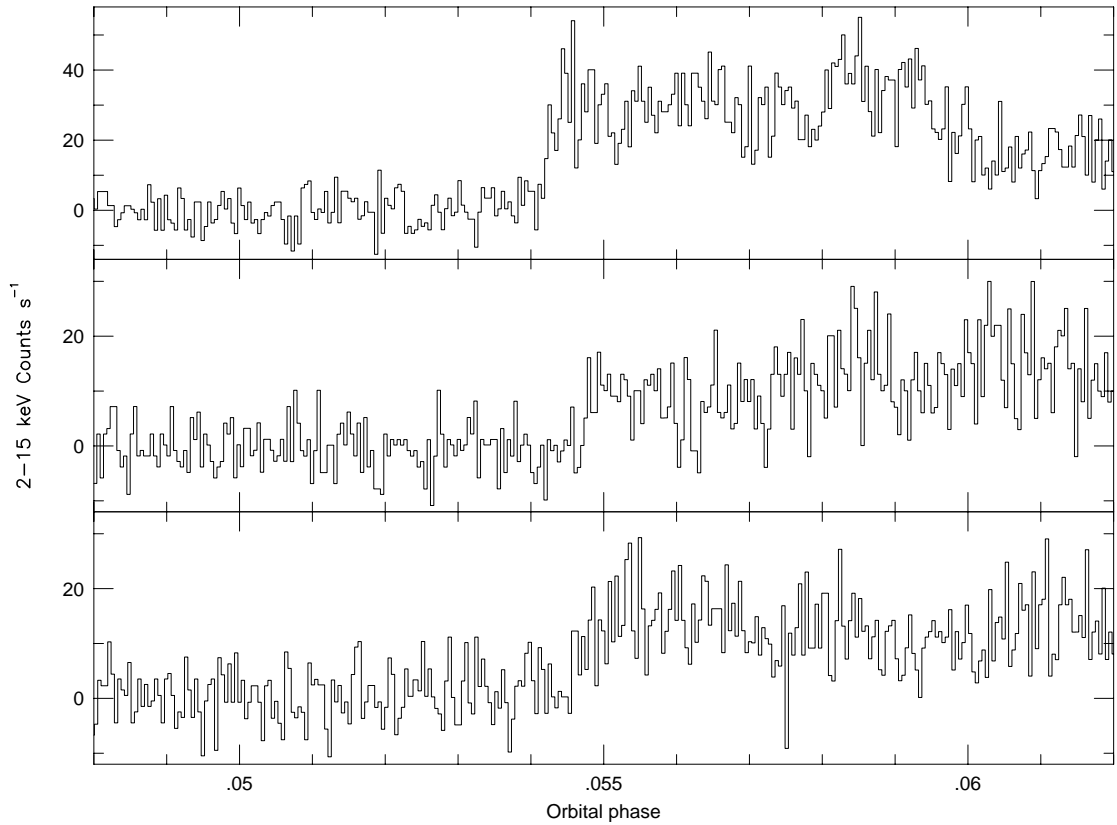}
\end{figure*}

\section{The eclipse egresses}
I first folded all 20 lightcurves on the orbital ephemeris of
Allan \etal\ (1996), and show the `averaged' egress in Fig.~3. 
To measure the times and length of egress I fitted the data with two constants,
one for the eclipsed light and one for the uneclipsed light, and joined these
by a straight line of variable slope. The best-fitting eclipse
egress lasted 25.8 s ($_{-1.8}^{+1.5}$ at 68 per cent confidence and 
$_{-4.2}^{+3.5}$ at 90 per cent confidence), while the phase of mid-egress was
0.0546(1).\footnote{ Warning: since I have no eclipse ingresses I can't
estimate the absolute orbital phase reliably, and so use the Allan \etal\
(1996) ephemeris throughout. However, comparison with the length of the {\it
Ginga\/} eclipse suggests that this ephemeris has accumulated a phase error of
0.0077, which should be subtracted from any phases to give a better estimate of
the absolute value.} The start of egress (3\up{rd}\ contact) is very well
defined but the end is rounded, introducing most of the uncertainty.  In fact
the star appears \sqig2 counts s\up{--1}\ brighter immediately after  egress
and also slightly brighter just before egress (although the latter is not
statistically significant). The post-egress brightening (if it is  flickering)
would  tend to bias the egress measurement to a greater duration. If the
brightenings are not flickering they might be explainable by X-rays scattered
in the corona of the secondary star into the line of sight. 

The egress in Fig.~3 is the average over spin phase, so the length of egress
depends on the size of the tracks over which accretion can occur through the
spin cycle, rather than on the size of the accretion region at any one time. To
investigate that we need to look at the individual eclipses. Fig.~4 shows three
of the clearest individual egresses (in general the others have lower S/N, or
are harder to interpret owing to greater source variability just after, and
presumably in, egress). It is immediately clear that the egress occurs at
different phases in the different eclipses. Thus, let's consider the phase
shifts we expect in the simplest case.

Fig.~5 illustrates the white dwarf emerging from eclipse. An accretion region
(spot) moves along a track of constant latitude as the white dwarf rotates. At
such a high inclination the accretion region will only be in view for half the
spin cycle, spending the other half behind the white dwarf. The fact that the
quiescent spin pulse has a low amplitude (Sections 2 \&\ 6) implies that we
must see accretion regions at both poles of a roughly symmetric dipolar field,
so that the appearance of one pole compensates for the disappearance of the
other. Thus the accretion region at the upper pole comes over the limb into
view at spin phase 0.25, travels across the face of the white dwarf, and
disappears at phase 0.75; also at phase 0.75, the accretion region at the lower
pole comes into view and travels across the white dwarf, before disappearing at
phase 0.25. Of course the chosen phase zero point is arbitrary, but reflects
the convention that phase zero is taken as the maximum of the spin pulse, and
my prejudice that in IPs this typically occurs when the upper 
pole points away from the observer.

The above assumes that the height of the accretion regions above the white
dwarf surface is negligible, a point to which I return in Section~5. Fig.~5
also contains two simplifications which are unimportant compared to other
uncertainties, but which make the egress times amenable to simple geometry. I
have assumed that the limb of the secondary is straight (at a constant angle
from the vertical of $\theta$) rather than curved; and similarly that the
inclination is sufficiently high that a line of constant latitude projected
onto the plane of the sky can be taken as straight, rather than slightly
curved.

To measure the egress times in the individual lightcurves I first tried a
generalisation of the fitting procedure described for the averaged profile.
This modelled the un-eclipsed light with sinusoids at the spin period and its
first harmonic, and multiplied this by the linear cutoff during egress.
However, the flickering and cycle-to-cycle variability in the lightcurves is
large, and this can lead to a bias when the fitting code lengthens the egress to
model brightenings occurring after egress. I therefore turned to a different
method of measuring 3\up{rd}\ contact, which proved to be more robust. This
simply stepped through the lightcurve starting in eclipse, calculated the
\chisq\ about zero flux of the data so far, and recorded the time at which the
\chisq\ rises significantly. By this method I arrived at the 19 times of
3\up{rd}\ contact plotted in Fig.~6 (one eclipse was discarded owing to a data 
drop-out during egress).

Fig.~6 shows a clear trend to later eclipse times with later spin phase
(as expected if the white dwarf rotates in the same sense as the binary, as it 
almost certainly must). The earliest eclipse times occur when
the upper pole has just come over the limb, and, as discussed above, I have
adjusted the phasing in Fig.~6 so that this occurs at spin phase 
0.25. Half a cycle later, when the upper pole disappears and the lower 
pole appears, the contact times jump earlier. The dashed lines illustrate the 
expected change in contact phase over the spin cycle, following the geometry
in Fig.~5 (the track for each pole is sinusoidal since it is the transverse
component of the motion which affects the phase of contact). The two free 
parameters affecting the tracks (their total phase span, corresponding to $l$ 
in Fig.~5, and their separation at phase 0.75, corresponding to $x$) were
adjusted to match the data. The total phase span of 3\up{rd}\ contact times is
$22\pm2$ s, compatible within the errors with the egress length of the averaged
lightcurve, which provides a check on the correctness of the procedure so far. 

The one abberant point in Fig.~6 (plotted as a triangle) occurred at the peak
of the outburst, and is discussed further in Hellier \etal\ (1997). The
second-brightest outburst observation was the point omitted because of a data
drop-out. The remaining outburst points all occur close to quiescent points
having the same spin phase, so the plot would not change significantly if only
quiescent points were included.

\begin{figure}\vspace{10cm}   % Fig 5
\includegraphics{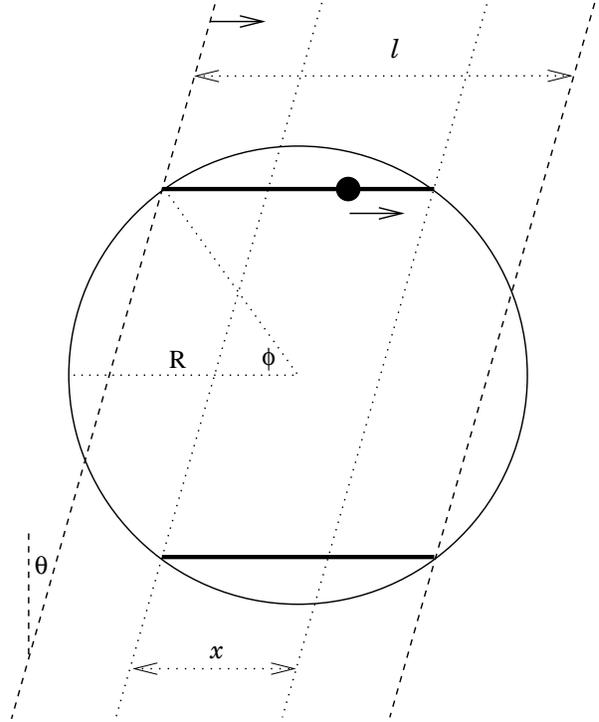}
\caption{An illustration of the egress of the white dwarf as the secondary star
limb moves from left to right.  It assumes that X-ray emission comes only from
regions of fixed latitude $\phi$. $\theta$ is the angle of the secondary star
limb, $R$ the radius of the white dwarf and $l$ the motion of the limb between
3\up{rd}\ and 4\up{th}\ contact in spin-averaged data.}
\end{figure}

\begin{figure}\vspace{8.5cm}   % Fig 6
\caption{The orbital phases of 3\up{rd}\ contact for each egress, against the 
spin phase at which 3\up{rd}\ contact occurred. The unfilled symbols indicate
lightcurves which are less reliable because of a lower count rate. The
discrepant point plotted as a triangle occurred at the peak of the outburst.
The uncertainties in orbital phase are typically 5\pten{-5}, or up to
2\pten{-4}\ for unfilled points. The dashed lines show the expected track of a
point on  the white dwarf (see text). The total phase span of the dashed lines 
corresponds to the length $l$ in Fig.~5, while their separation at phase 0.75
corresponds to $x$.}
\includegraphics{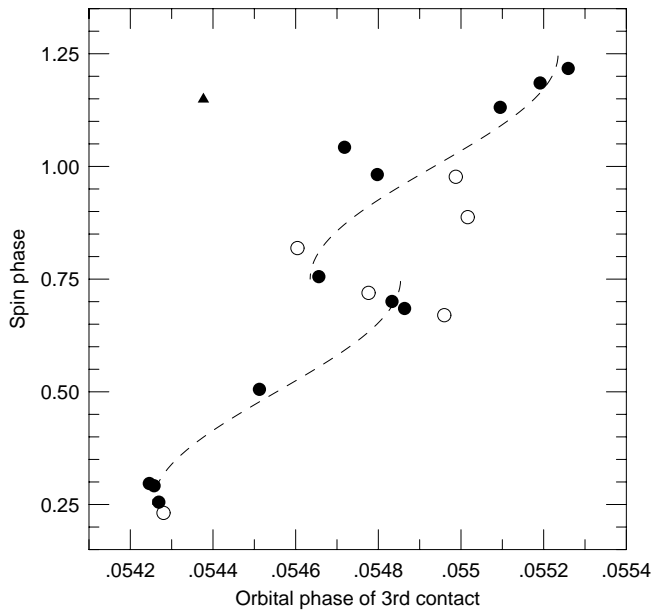}
\end{figure}

\section{The location of the accretion regions}
In order to translate the egress times in Fig.~6 to locations on the white
dwarf we need to review the geometry of the binary. From ellipsoidal modelling
of the IR lightcurve Allan \etal\ (1996) limit the mass ratio to 
0.4\,$<$\,$q$\,$<$0.7 and the inclination to 80\deg\,$<$\,$i$\,$<$87.  We can
also use the fact that the secondary fills the Roche lobe of the 6.06-h
binary, together with a mass--radius relation (e.g.~Patterson 1984; Warner 
1995), to estimate the secondary mass as 0.62 \msun. Together with the limits on $q$
this implies a white dwarf mass of \sqig1 \msun.

As mentioned above, though, both the distribution of points in Fig.~6 and the
fact that the spin pulse has a low amplitude suggest that we are seeing both
poles of the white dwarf. This further restricts the inclination, as shown by
the following. Accretion will be inhibited by a centrifugal barrier  if the
velocity of the field lines at the point of disc disruption exceeds the local
Keplerian velocity. Assuming that they are the same, the 206-s spin period and
1 \msun\ white dwarf mass locate the inner radius of the disc at
5.2\pten{9}\,cm, compared to a white dwarf radius of 5.6\pten{8}\,cm. This
implies that the inclination must be no higher than 84\deg\ in order to see the
bottom of the white dwarf, even assuming that the disc has zero thickness at
the disruption point. 

I therefore proceed with the range 80\deg\,$<$\,$i$\,$<$84\deg.  The length of
the X-ray eclipse (2050 s; Kamata \etal\ 1991; Allan \etal\ 1996) then fixes
the mass ratio to 0.48\,$<$\,$q$\,$<$0.68, using figure~2 of Horne (1985). 
For these values of $q$ and $i$ the angle of the secondary star limb
at the white dwarf ($\theta$ in Fig.~5) ranges from 20\deg\ ($i$=84\deg)
to 30\deg\ ($i$=80\deg).

From the geometry illustrated in Fig.~5 one can show that 
\[ \frac{l}{2R} = \frac{\cos(\theta-\phi)}{\cos \theta} \]
and 
\[ \frac{x}{2R} = \frac{\cos(\theta+\phi)}{\cos \theta} \]
and hence
\[ \frac{x}{l} = \frac{\cos(\theta+\phi)}{\cos(\theta-\phi)}. \]

The ratio $x/l$ can be gained from Fig.~6; a formal fit to the plotted
tracks yields $x/l = 0.22 \pm 0.06$. Hence, including the uncertainty
in $\theta$, the latitude of the accretion regions, $\phi$, lies in the
range 44\deg\ to 63\deg. Further, the ratio $R/l$ will be 0.45--0.64 (at
the lower end of the range for $\theta = 30$ and the upper end for $\theta
= 20$).

To translate $l = 25^{+1.5}_{-1.8}$ s into a length we need the orbital speed
of the secondary star with respect to the white dwarf. From Kepler's law, \[
v^{3} = \frac{2\pi G (M_{1}+M_{2})}{P}, \] where $v$ is the velocity and $P$ the
orbital period. Assuming M\lo{2} = 0.62 \msun, the velocity is between 390
\kmps\ ($q$\,=\,0.68) and 420 \kmps\ ($q$\,=\,0.48). Hence we finally arrive at
a white dwarf radius in the range 4.3--7.3 \pten{8}\,cm. Using the Hamada \&\
Salpeter (1961) mass-radius relation for a white dwarf, but increasing the
radius by 5 per cent following the work of Koester \&\ Schoenberner (1986),
this translates to a mass in the range 0.78--1.17 \msun. 

The limits on the white dwarf mass are not much different from those given by
the range of $q$ and the assumed secondary star mass-radius relation
(0.91--1.29 \msun); however the consistency does suggest that the procedure so
far, and the  interpretation of Fig.~6, is essentially correct. Thus, to sum
up,  the accretion regions are seen at both poles, lying at a latitude between
44\deg\ and 63\deg, on a white dwarf whose diameter is a length translating to 
between 22 and 35 s of secondary star motion.
 
\section{The size of the accretion region}
Again, before seeing what the data say about the size and shape of the
accretion region, let's review the standard picture of what we expect. The 
magnetic field will pick up material from the inner edge of the disc and
channel it along field lines onto the white dwarf. The range of azimuth in the
disc translates, with sufficient accuracy for present purposes, to a ring of 
constant magnetic latitude around the magnetic pole (see, e.g.,~Rosen \etal\ 
1988; Kim \&\ Beuermann 1995). If the disc is disrupted at a radius of \sqig9
R\lo{wd}, as suggested in Section~4, the magnetic colatitude of this ring,
$\epsilon$, will be \sqig19\deg. The magnetic axis will be offset from the
orbital plane by an angle $\delta$ (otherwise there would be no spin pulse) so
we expect a variation in accretion rate with azimuth such that most accretion
arrives from the azimuth to which the magnetic pole points, translating to a
location on the ring around the magnetic pole directly opposite the spin axis
(Fig.~7). The accretion rate would fall off in either direction along the ring,
producing arc shaped accretion regions extended along lines of constant
magnetic latitude. From the previous section we have the constraint that
$\epsilon + \delta = 90 - \phi$, where $\phi$ lies in the range 44--63\deg.
With $\epsilon \sim 19^{\circ}$ this implies $8^{\circ} < \delta < 27^{\circ}$.

\begin{figure}\vspace{10cm}   % Fig 7
\includegraphics{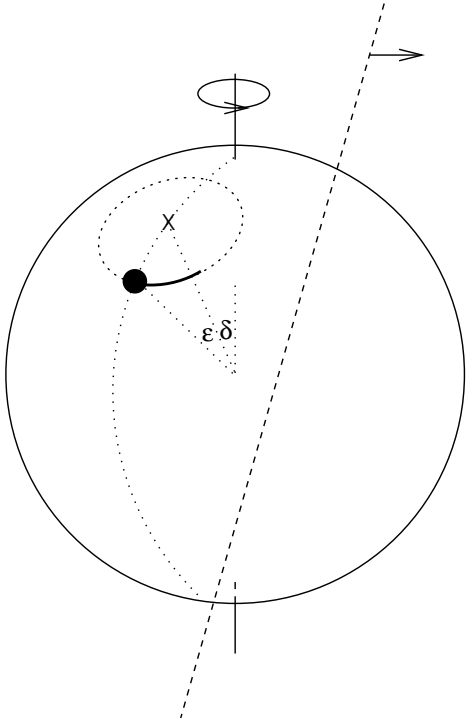}
\caption{An illustration of the expected accretion regions in a magnetic
accretor. The magnetic pole, {\sf X}, is offset from the spin axis by an
angle $\delta$. The accretion regions lie approximately on a circle of 
constant magnetic latitude, at a further angle $\epsilon$ from the magnetic
pole. The highest accretion rate is expected furthest from the spin axis.
As the white dwarf spins this point moves along a line of constant latitude,
as shown in Fig.~5.}
\end{figure}

\begin{figure*}\vspace{8cm}   % Fig 8
\caption{The average eclipse egress using an empirical phase correction 
(see text). Each bin lasts 1 s.}
\includegraphics{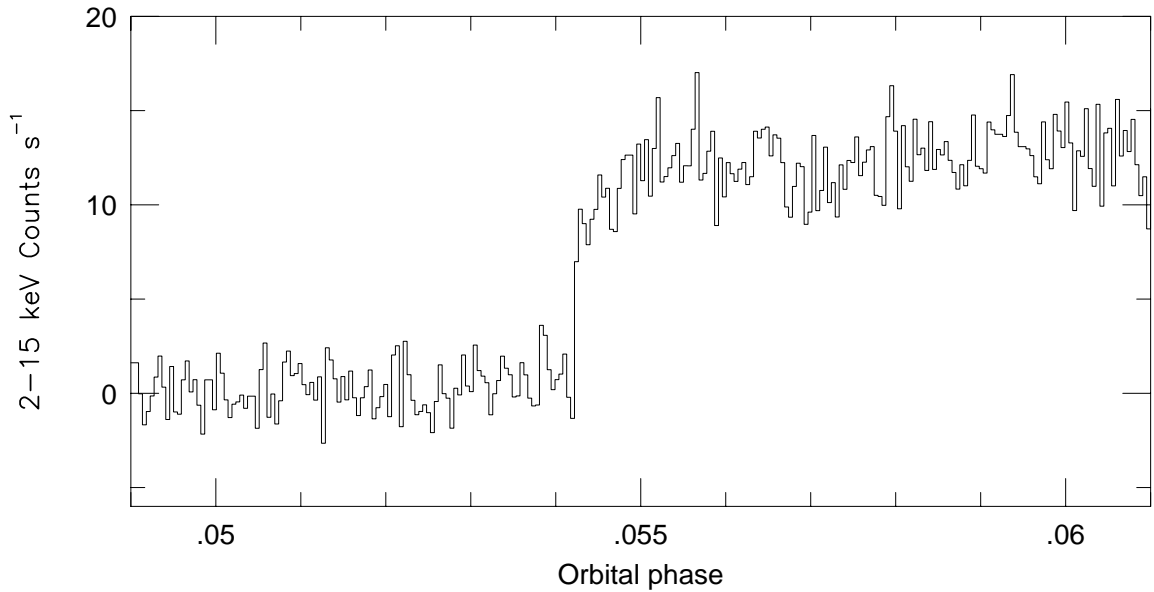}
\end{figure*}

Turning to the data, one could investigate the size of the accretion region
either by looking at individual eclipses, or by summing the eclipses, shifting
them in phase to account for their different locations on the white dwarf. The
first method suffers from low S/N while the second is risky, since if the phase
adjustments are wrong the result will be artificially smeared out. 

Pursuing the second method, Fig.~8 shows the egress formed by combining the
eclipses while subtracting from each the phase of its location on Fig.~6 (for
this and the following results I omitted the two eclipses at the outburst
peak). Two thirds of the flux emerges in $<$2 s while the remaining third
emerges more slowly, so that the whole egress lasts \sqig20 s.  Could the
$<$2 s egress be an artefact of lining up flickers in individual eclipses?
To some extent it could, but if the flickers were photon noise one could
only line up individual bins, and would expect the subsequent bins to revert
to the average trend. The fact that there is no decrease after the rapid
egress shows that it cannot be photon noise. Similarly the rapid egress
is unlikely to be an artefact of source flickering, since this would
imply that 18 eclipses had conspired to show excess light through flickering
at exactly the moment the accretion regions began to be uncovered. Further,
under the hypothesis that the true egress were slower, most of the accretion
region  would still be hidden at the beginning of egress. It is implausible
that the small region visible could, on its own, produce flickers of
2/3\up{rds}\ the intensity of the entire region. 

The slower, \sqig 20 s, egress is more plausibly an artefact,  since if any
of the eclipses had the wrong phase correction, again possibly through
flickering, a smeared egress would  result. Again, though, flickering is least
likely to be a factor in measuring 3\up{rd}\ contact, and a careful examination
suggested that the slower section of the egress is real. 

In contrast to the `empirically' corrected egress in Fig.~8, Fig.~9 shows the 
combined egress, this time subtracting the phase of the dashed lines in Fig.~6,
appropriate for the spin phase of the egress. For comparison the uncorrected
egress from Fig.~3 and the empirically corrected egress are shown again. The
egress using the calculated corrections is faster than the uncorrected egress,
with 2/3\up{rds}\ of the flux emerging in \sqig5 s, but it is slower than the
empirically corrected egress.

My interpretation of Figs.~8 \&\ 9 is as follows: the majority of the X-rays
(\sqig 2/3\up{rds}) emerge, at any one time, from a very small region
($<2$\,s), located at the trailing edge of the accretion region. The remainder
of the flux arises from a much more extended region which leads the small,
bright region, and produces the slower part of the egress.

Further, the location of the bright region is not determined solely by spin
phase, but can deviate from its spin-averaged location by up to \sqig 2 s, so
that lining up the eclipses using spin phase only, rather than the empirical
position, smears the bright spot over a larger region. 

Having interpreted the summed egress from all the eclipses, I have checked for
consistency with the individual egresses. Because of the low S/N these only
give upper limits on the egress times. The three eclipses in Fig.~4 give upper
limits of (from top to bottom) 9, 4 \&\ 5 s (68 per cent confidence) or 11, 10,
15 s (90 per cent confidence). The other eclipses gave weaker limits. Thus all
the individual eclipses are consistent with a rapid egress in 2-s, but their
low S/N limits their usefulness. The slower part of the egress is not
traceable in the individual lightcurves. 

\subsection{Complicating factors}
There are several complications to be considered before translating the 
measured times into sizes on the white dwarf. 

(i) If the secondary limb is not a sharp edge the  transitions will be
lengthened. This, though, simply makes the above upper limits firmer. The
egress time of $<$2 s implies that the secondary edge is sharp to 800 km, or
$\Delta r/r$ \sqiglt 2\pten{-3}. The hard X-rays will be stopped by a column
$>10^{24}$ cm\mintwo, while the path length through the secondary limb will be
\sqig10\up{10}\ cm, implying that the secondary edge occurs at a density of
\sqig10\up{14}\ cm\up{--3}. This will be located in the chromosphere at a point
where the density increases by an order of magnitude over a vertical distance
of 300 km (e.g.~Withbroe \&\ Noyes 1977). Thus, fuzziness of the secondary star
becomes important at about the 1 s level.

A similar consideration is that if the atmosphere of the secondary changes,
because of sunspots for instance, so that the location of the edge varies by 
$\Delta r/r$ \sqiglt  2\pten{-3}, it could explain some of the apparent phase 
wandering seen when lining up the eclipses using the spin phase only.

(ii) Many of the X-ray photons will be scattered into a halo as they pass
through the molecular cloud Lynds 1457. Mauche \&\ Gorenstein (1986) 
find that for GX\,13+1 and Cyg~X-3, with columns of 5\pten{22}\ and 3\pten{22}\
cm\mintwo\ respectively, the halo contains 0.19 and 0.33 of the total
intensity at an energy of 2.4 \kev. XY~Ari has a somewhat larger column of 
8\pten{22}\ cm\mintwo; however the 2--15 \kev\ pass band will give a higher
flux-weighted mean energy, $E$, of 6 \kev, and the scattering cross-section
decreases as $E^{-2}$. Thus we would expect 5--15 per cent of the photons 
to be scattered. 

The time delay caused by the longer path length can be estimated using
(e.g.~Bode \etal\ 1985)
\[ \Delta t \approx 0.22 \left(\frac{a}{0.25 \mu{\rm m}}\right)^{-2}
\left(\frac{E}{1 {\rm \kev}}\right)^{-2}
\left(\frac{D}{2.5 {\rm kpc}}\right) {\rm days}.\] 
For a grain size, $a$, of 0.1 $\mu$m, an energy of 6 \kev, and a 
source distance, $D$, of 300 pc, this gives a delay of \sqig 400 s.

The two estimates suggest that the halo is somewhat too weak and varying
too slowly to cause the slower part of the observed egress (one third of
the flux emerging in 20 s), although it could become important in higher
S/N data. Halo photons could also explain the excess of photons in
mid-eclipse above the estimated background (Section~2).

(iii) The calculations so far have assumed that the shock height above the
white dwarf surface can be neglected. Standard theory (e.g.~Frank, King \&\
Raine 1992) yields an upper limit to the shock height given by \[ D \sim 9
\times 10^{9} \dot{M}_{17}^{-1} f M_{1}^{3/2} R_{9}^{1/2} {\rm cm} \] where
$\dot{M}_{17}^{-1}$ is the accretion rate in $10^{17}$ g s\minone, which can be
taken as \sqig 1 for an IP above the period gap (e.g.~Patterson
1994); $M_{1}$ is the primary mass in solar units, which from above is \sqig1;
$R_{9}$ is the white dwarf radius in $10^{9}$ cm, which is \sqig0.55; and $f$
is the accretion area as a fraction of the white dwarf. 

Assuming that the bright spot is circular with a diameter of 2 s, and that the
white dwarf is spherical with a diameter of at least 22 s, we have $f <
0.002$. This gives $D < 0.02 R_{\rm wd}$ and justifies the assumption of
negligible height. However, for the extended region, $f$ could be 0.1, and
since the extended region emits \sqig1/3\up{rd}\ of the flux, its share of
$\dot{M}$ could be \sqig 1/6\up{th}\ per pole, implying $D \sim R_{\rm wd}$.
This means that the extended region could have a large height, rather than a
large area, even though the emission from an accretion column will emerge
largely from the higher density base of the column. However, the fact that none
of the slower egress precedes the rapid egress suggests that it does not have a
significant height, since this would cause it to egress earlier when the upper
pole is emerging over the limb.

(iv) The aspect of the accretion region presented to the white dwarf limb can
change. The egress, of course, is only sensitive to the extent of the region in
a direction perpendicular to the limb.  Fortunately, if we suppose that the
accretion region's shape is fixed relative to the magnetic field, the
orientation of the region will change as the white dwarf spins. Thus, since the
summed egress samples 18 different spin phases, the result will approximate to
an averaged cut across the region, especially since the egresses sample both
poles, which will also have different orientations. Nevertheless, this effect
should be borne in mind and a correction factor, either reducing or increasing
the size of the region, could be applied given greater knowledge of the shape
of the region. 

\begin{figure}\vspace{9cm}   % Fig 9
\caption{The average eclipse egress using a phase correction calculated from
the spin phase alone (see text). Each bin lasts 1 s. For comparison the
dashed line shows the empirically corrected egress and the dotted line the
uncorrected egress.}
\includegraphics{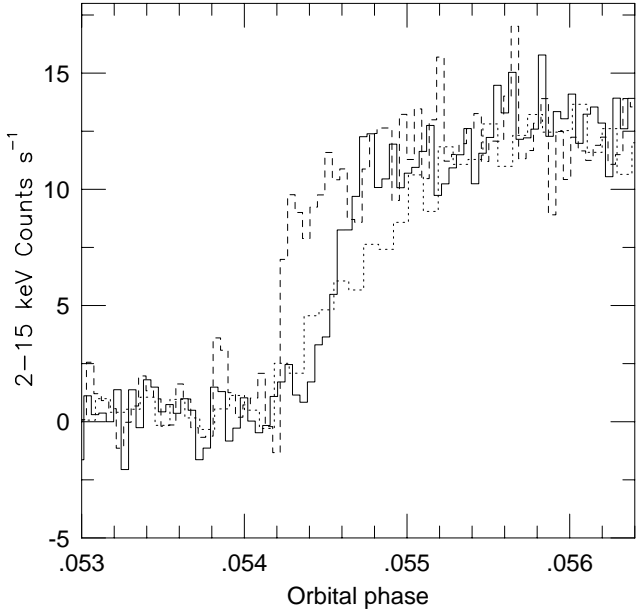}
\end{figure}

It is also possible, in principle, that the rapid and slower parts of the 
egress arise from one region which is extended into a long ribbon, giving both
slow and rapid egresses depending on its aspect. However, since the most likely
extension is along lines of constant magnetic latitude, this would lead to
longer egresses soon after white dwarf 3\up{rd}\ contact and shortly before 
white dwarf 4\up{th}\ contact, with shorter egresses in between (see Fig.~7). 
The data, though, show no such effect, with rapid egresses occurring at all
spin phases, suggesting that the small bright spot is real.  Note, though, that
this  conclusion would be more robust if the individual egresses had a higher
S/N. In particular the slower egress is only discernable in the summed egress
and can't be investigated as a function of spin phase.

A similar complication is that if the accretion regions have a negligible
shock height they will appear foreshortened when on the white dwarf limb, in a
direction perpendicular to the limb. However, while the angle between the
direction of foreshortening and the direction perpendicular to the secondary 
limb can be low during the appearance of the upper pole  ($\phi - \theta$, in
the nomenclature of Fig.~5) it will be larger at the disappearance of the pole
($\phi+\theta$; and similarly in reverse for the lower pole). Thus, when
averaged over all egresses, including many not near the limb, foreshortening
will have a minor effect.

(v) The accretion regions are themselves moving with the white dwarf rotation.
The white dwarf radius of 4.3--7.0 \pten{8}\,cm and spin period of 206 s imply
an equatorial surface motion of 130--210 \kmps. For an accretion region  at the
equator in the middle of the white dwarf the effective speed of the secondary 
is reduced by this amount, although the effect dies away sinusoidally in 
moving towards the pole or towards the limb, so that again the effect on the
averaged egress will be  minor. Since the correction to the averaged egress for
this effect will be in the opposite sense to the effect of foreshortening, and
since neither can be calculated securely given the uncertainties in the
locations of the regions on the white dwarf, it is easier to make neither
correction. 

\subsection{The value of $f$}
Bearing in mind the above complications, let's move to the value of $f$. The
rapid egress of 2/3\up{rds}\  of the X-ray flux occurs in $<$2 s. Assuming a
circular region with this value as the diameter is also likely to be an upper
limit, since the accretion region is likely to be arc-shaped and so have a
smaller area. Thus,  taking the smallest white dwarf diameter of 22 s (to again
give an upper limit), we obtain $f < 0.002$ for each pole. This is the area
over which the bulk of the accretion is occurring at any one time; $f_{\rm
acc}$ in the terminology of King (1995).

Recall, though, that the accretion spot appears to wander from its mean
position, so that 2/3\up{rds}\ of the flux emerges in 5 s in the spin-phase
corrected egress. This leads to $f_{\rm zone} < 0.01$ per pole, for the 
fractional area over which accretion can occur. Again this is likely to be an
upper limit since, whatever mechanism causes the wandering, it is likely to 
occur preferentially along either magnetic latitude or longitude (and we
can't rule out changes in the secondary star as the cause).

Finally, there is the remaining 1/3\up{rd}\ of the flux which appears to emerge
from a more extended region, having an egress time comparable with the white
dwarf radius (although I repeat that higher S/N data are required to confirm
its existence, since it could be an artifact of incorrect phase correction).
It is also harder to translate this into a value for $f$. Firstly, it is hard to
disentangle a higher shock height from a larger $f$, although as remarked
previously, the fact that none of the slower egress appears to precede third
contact with the white dwarf suggests that a higher shock is not an important
factor. Secondly, the most likely explanation of an egress time comparable with
the size of the white dwarf is accretion regions visible at both magnetic poles 
simultaneously, since they will be separated by a white dwarf diameter.
If there is any significant extension of the accretion arcs then parts of both
will be visible when they are on the limb, and the greater their extent the
greater the range of spin phases over which parts of both accretion regions are
visible. However, it is also possible that the slower egress does imply a
greatly extended emission region. If so, the summed egress gives knowledge of
one dimension, but not of the others. If we assume $f \sim 0.1$ we find an
estimated brightness, or flux per area, of 1/100\up{th}\ that of the bright
spot. Distinguishing between these possibilities will need higher S/N data
capable of tracing the slower egress in individual eclipses.

\section{The spin pulse}
I show the quiescent data folded on the spin period in Fig.~10, together with
the 6--15/2--5 \kev\ hardness ratio (the spin pulse in outburst is discussed 
in Hellier \etal\ 1997). Both the double-peaked profile and the hardness ratio
changes are similar to those in the {\it Ginga\/} data reported by Kamata \&\
Koyama (1993).  The phasing in Fig.~10 corresponds to that in Fig.~6, so that
the upper pole emerges over the white dwarf limb at phase 0.25.

\begin{figure}\vspace{13cm}   % Fig 10
\caption{The data from all 20 observations folded on the spin period,
together with the 6--15/2--5 \kev\ hardness ratio. The phasing is the
same as in Fig.~6, and is chosen so that the upper pole appears over the
white dwarf limb at phase 0.25. Phase 1 occurs at TDB 245\,0277.40382}
\includegraphics{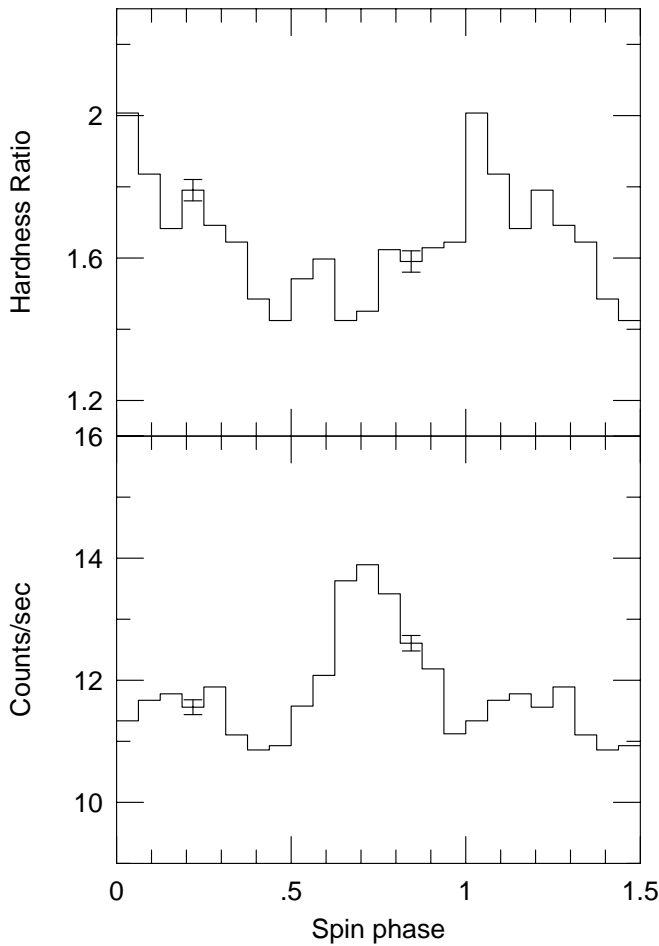}
\end{figure}

As previously mentioned, the amplitude of the spin pulse is much lower than
in IPs such as AO~Psc, which has an \sqig 80 per cent pulse
(e.g.~Hellier \etal\ 1996). Thus the appearance of one pole over the
limb must compensate for the disappearance of the other pole. In a
symmetric situation with negligible shock heights this would lead to no
spin pulse. The low amplitude, double-peaked pulse in XY~Ari can then be
attributed to asymmetries between the poles, or the effect of a non-negligible
shock height for at least some of the X-ray flux. 

At phase 0.25 in Fig.~10 the flux starts to decline. Since this is when the
lower pole disappears and the upper pole appears, it suggests that the lower
pole is slightly brighter. The maximum flux is seen half a cycle later, near
phase 0.75, when the lower pole would be appearing and the upper pole 
disappearing. It is possible that they aren't directly opposite each other so
that more of both regions is visible for a short time, producing extra flux.
Being more specific is hard without a greater knowledge of the shape and extent
of the accretion regions.

The hardness ratio is greater in the phase range 0.75--0.25, when the flux
comes from the lower pole. Normally, this would argue against the accretion
curtain model for the spin modulation, which proposes that greater absorption,
and hence a harder spectrum, would be expected when the upper pole points
towards us (e.g.~Hellier \etal\ 1996). However, as shown in Section~4, the
inclination of XY~Ari is so high that the line of sight to the lower pole skims
over the inner edge of the disc. It is therefore likely that this line of sight
will encounter far more absorption than the line of sight to the upper pole,
which can explain the different behaviour.

\section{Discussion}
The essential result of this paper is that the majority of the X-ray flux
emerges, at any one time, from regions covering $<$\,0.002 of the white
dwarf surface at each pole. These regions, however, wander in
location, covering an area of $<$\,0.01 of the white dwarf. Further, there
is evidence that either we can often see some flux from both
poles simultaneously, or that a minority of the X-ray flux emerges from a 
much more extended region, probably preceding the bright region in phase.
Note that these results derive from quiescent observations. In Hellier 
\etal\ (1997) we deduce that the accretion regions become much more extended
in outburst.

There are only a few other estimates of the accretion area in IPs for
comparison. Hellier (1993) argued that the upper pole in FO~Aqr must have
$f$\sqig0.001, but this was from a model-dependent interpretation of the spin
pulse. Also, some of the IPs discovered with {\it Rosat\/} show strong
black-body emission, presumably from the heated pole-cap.  Haberl \etal\ (1995)
use this to estimate $f \sim 10^{-5}$ for RX\,0558+53 for a distance of 300 pc,
although as usual the distance is suspect. Similarly, the polarised light from
PQ~Gem allows an estimate of the area emitting cyclotron radiation, and Vath,
Chanmugham \&\ Frank (1996)  model the emission to find $f_{\rm cyc} \sim
0.006$. Further, values of $f$ \sqig 0.01--0.001 are expected using current
(but highly uncertain) theoretical expectations for the magnetic disruption of
a disc (e.g.~Rosen 1992).

\begin{table}\caption{System parameters of XY~Ari}
\begin{tabular}{lcl}
Orbital period & 21832.990(5) & secs \\
Spin period    & 206.298(5)    & secs \\
Inclination, $i$    & 80--84        & degrees \\
Mass ratio, $q$ & 0.48--0.68 \\
White dwarf radius & 4.3--7.0 \pten{8} & cm \\
White dwarf mass & 0.91--1.17 & \msun \\
Red dwarf mass & \sqig0.62     &  \msun \\
Inner disc radius & \sqig 9    & $R$\lo{wd} \\
Latitude of accretion, $\phi$ & 44--63 & degrees \\
Magnetic colatitude of accretion, $\epsilon$ & 19 & degrees \\
Magnetic dipole offset, $\delta$ & 8--27 & degrees \\
Accretion area, $f_{\rm acc}$ & $<0.002$ \\
Range of accretion area, $f_{\rm zone}$ & $< 0.01$ \\
\end{tabular}\end{table}

In AM~Her stars many more estimates are available from the blackbody radiation,
from cyclotron emission, and from eclipses. Typical values are $f_{\rm cyc}$
\sqig 3\pten{-4}, $f_{\rm bb} \sim 10^{-5} - 10^{-6}$ (e.g. Stockman 1995) and,
from eclipses of UZ~For, $f_{\rm acc} < 0.005$ (Bailey \&\ Cropper 1991).

Thus all values are consistent and suggest that polecap areas have $f$ of order
$10^{-3}$ in IPs and are even smaller in AM Her stars. The
larger $f$ in IPs is expected since an accretion disc can feed
accretion over a larger range of azimuth than a stream will.

One consequence of this is that occultation of such regions will produce sharp
features in the X-ray spin-pulse profiles (barring special grazing geometries)
and so can explain features such as the `notch' in the spin pulse of FO~Aqr.
The much broader quasi-sinusoidal modulations typical of IPs 
must be caused by absorption, which fits with their characteristic greater
depth at lower energies. The high spectral resolution of {\it ASCA\/} data
confirms this at least in the case of AO~Psc (Hellier \etal\ 1996). 

The small accretion areas also imply very high densities in the accretion
column. Rosen (1992) shows that this will produce substantial electron
scattering, and so, by beaming the X-ray flux from the column, spin-modulate 
the X-rays even at high energies. This, in turn, implies that the material must
be highly ionized (as expected in the post-shock region) or otherwise an
inefficient absorber, in order for any low energy X-ray flux to be seen at 
spin minimum.

Despite the above, though, absorption seems to contribute little to the spin
pulse in XY~Ari. Instead, the low-amplitude, complex, double-peaked  pulse
appears to result from asymmetries between the accretion regions at the two
magnetic poles, so that occultation effects do not entirely cancel. Why the
difference between it and AO~Psc? One reason is that XY~Ari is already hidden
behind a column of $10^{23}$ H atom cm\mintwo, provided by the molecular cloud,
so that changes in more tenuous absorbers will have little effect. In AO~Psc we
found that the absorber consisted of several phases: a dense 2\pten{23}\
column, associated with the flow very near the white dwarf, immediately before
the accretion shock; and a less dense leaky absorber of 2\pten{22}, associated
with parts of the accretion flow far from the white dwarf, possibly where the
magnetically channelled material rises above the plane of the accretion disc.

Thus, a possible resolution involves the higher inclination of XY~Ari. Looking
virtually along the plane of the binary, with the accretion columns falling
onto the magnetic poles of the binary at an angle of between 43\deg\ and
63\deg\ to the plane, the denser pre-shock material would never obscure the
line of sight. Thus the only absorption in XY~Ari would result from the
material rising out of the plane near the magnetosphere, and if this had a
similar density to that in AO~Psc it would be unnoticeable given the greater
column of the molecular cloud. In the lower inclination AO~Psc ($i$
\sqig\,60\deg) the pre-shock material would cross the line of sight, and so
cause the very deep, absorption-induced spin pulse.

It is notable that the other eclipsing IP, EX~Hya, also has a
spin pulse showing very little sign of absorption (e.g.~Ishida, Mukai \&\
Osborne 1994; Allan, Hellier \&\ Beardmore 1997). Probably the high inclination
reduces the effect of absorption in this star also. [Ironically, despite the
fact that the accretion curtain model was first developed for EX~Hya, it now
appears that it applies far better to stars such as AO~Psc and does not explain
the X-ray behaviour of EX~Hya.]

Although the data show that the accretion area, $f_{\rm acc}$, is $<0.002$
at any one time, the area over which the accretion wanders, $f_{\rm zone}$,
is larger at $<0.01$. Since the eclipses are distributed over a month and
are typically separated by 1 day, we can't tell whether the wandering occurs 
from spin cycle to spin cycle, or on a timescale of days. Presumably the
wandering reflects fluctuations in the connection of the field lines
onto the inner disc, and if this takes place on short timescales it might
explain much of the flickering in this and other IPs.

The existence of a slower section of egress in addition to the rapid egress
is a sign of structure in the accretion region. If it is caused by viewing 
emission from both poles simultaneously it implies that accretion occurs
along an extended arc around each magnetic pole, and hence that 
each pole receives at least some material from a range of azimuths.
The different angle of the field lines to the disc at different azimuths
would then modulate the rate of accretion with azimuth to produce
much more intense emission at one location on the ring, and thus the 
bright spot responsible for the rapid egress.

If, instead, the accretion regions at the pole disappear completely behind
the white dwarf, the slower egress implies a large but faint extension to the 
bright spot. The current egresses have too low a S/N to trace this feature
in individual egresses, so further investigation will have to await future
satellites. However, the fact that the slower egress occurs largely after
the rapid egress implies that the faint extension leads the bright spot
on the white dwarf, suggesting that field lines ahead of direction in which
the magnetic axis points accrete preferentially compared to those behind.
Interestingly a similar asymmetry has recently been suggested for PQ~Gem,
based on X-ray data (Mason 1997), optical spectroscopy (Hellier 1997) and
polarimetry (Potter \etal\ 1997). Thus we seem to be accumulating the first
observational clues to the structure of the interaction between the magnetic
field and the inner disc.

\section{Conclusions}
(1) The majority of the X-ray flux emerges from eclipse in $< 2$ s, implying 
that the accretion polecaps have areas of $<0.002$ as a fraction of the
white dwarf surface.

\noindent (2) The accretion footprints are not fixed in position on the white
dwarf, but can wander over an area $<0.01$ of the white dwarf surface. 

\noindent (3) There is evidence that a minority of the flux arises from a much
larger area, or that we can always see part of the accretion regions at both
poles.

\noindent (4) The accretion regions are at a latitude in the range
43\deg--63\deg, which suggests a magnetic dipole offset of 8\deg--27\deg\
(constraints on other system parameters are listed in Table~1).

\noindent (5) We see roughly equal amounts of X-ray flux from both upper and
lower poles. There are, though, slight asymmetries so that the disappearance of
one pole over the white dwarf limb does not entirely compensate for the
appearance of the other pole. The net effect is a low-amplitude, complex spin
pulse.

\section*{Acknowledgments}
I thank Tim Naylor, Koji Mukai, Rob Jeffries and Nye Evans for valuable
discussions. I am also grateful to the {\it RXTE\/} team for their execution of
a difficult observation and for the assistance provided by the e-mail help
desks.

\end{document}